\documentclass[aps,prl,reprint,groupedaddress]{revtex4-1}

\usepackage{graphicx}
\usepackage{dcolumn}
\usepackage{bm}

\begin{document}

\title{Non-equilibrium steady structures of confined liquid crystals driven by a dynamic boundary}

\author{Rui-fen Zhang$^{1}$, Chun-lai Ren$^{1,\dag}$, Jia-wei Feng$^{1}$, and Yu-qiang Ma$^{1,\dag}$}

\affiliation{
$^1$ National Laboratory of Solid State Microstructures and Department of Physics, Collaborative Innovation Center of Advanced Microstructures,\\
          Nanjing University, Nanjing 210093, China\\
          $\dag$ E-mail: chunlair@nju.edu.cn (C. R.); myqiang@nju.edu.cn (Y. M.)
}
\date{\today}
\begin{abstract}
Steady structures originating from dynamic self-assembly have begun to show their advantages in new generation materials, and pose challenges to equilibrium self-assembly. In view of the important role of confinement in self-assembly, here, we propose a new type of confinement leading to dynamic steady structures, which opens a new window for the conventional confinement. In our model, we consider the self-assembly of ellipsoids in 2D circular confinement via the boundary performing periodically stretching and contracting oscillation. Langevin dynamics simulations reveal the achievement of non-equilibrium steady structures  under appropriate boundary motions, which are novel smectic structures with stable topological defects. Different from the confinement with a static boundary,  ellipsoids close to the boundary have variable orientations depending on the boundary motion. Order-order structural transitions, accompanied by the symmetry change and varied defect number, occur with the change of oscillating amplitude and/or frequency of the boundary. Slow and fast dynamics are distinguished according to whether structural rearrangements and energetic adjustment happen or not. The collective motion of confined ellipsoids, aroused by the work performed on the system, is the key factor determining both the structure and dynamics of the self-assembly. Our results not only achieve novel textures of circular confined liquid crystals, but also inspire us to reconsider the self-assembly within the living organisms.
\end{abstract}

\maketitle

\section{Introduction}\label{section1}
Self-assembly is widely used in creating complex artificial nanostructures and building novel nanomaterials and devices\cite{shevchenko2006structural,velev2009materials,singh2014self}. Recently, much attention has been paid to dynamic self-assembly, characterized by energy input and dissipation. This is an important way for self-organization in nature. In particular, the steady structures resulting from dynamic self-assembly have the ability to adapt to environmental changes, which could be good candidates for new materials\cite{glotzer2015assembly}. For example, it was reported that a nanoscale device for transmission of torque can be realized by colloids in 2D confinement driven by a rotating boundary\cite{williams2016transmission}, where the realization of controllable dynamic self-assembly structures of confined colloids is the key. Currently, to implement dynamic steady structures, stimuli are frequently used based on the idea of changeable effective interactions among building blocks, such as external fields\cite{grzybowski2000dynamic,martin2013driving,lowen2008colloidal}, light\cite{palacci2013living}, chemical reactions\cite{nedelec1997self}, and pH\cite{tagliazucchi2014dissipative}. However, new strategies are still very limited due to the lack of understanding the mechanisms of nonequilibrium self-assembly and the relationship between dynamics and structure.

The confinement provides an effective way to realize rich self-assembly structures. A representative example is confined liquid crystals (LCs)\cite{lopez2011frustrated,lin2011endotoxin,garlea2016finite}, where self-organized structures can be frustrated by the existence of boundaries leading to topological defects\cite{poulin1997novel,Zapotocky1999Particle,muvsevivc2006two,araki2011memory,Lavrentovich2010Nonlinear,Pishnyak2011Inelastic}.
In most confined LCs cases  where the particles are packed at high density and their size  is comparable to the confining space
 (i.e., the strong confinement conditions)\cite{garlea2016finite,confine1,confine2},  the confinement plays a significant role  in stabilizing  and manipulating the topological defects\cite{garlea2016finite,manyuhina2015viral,press1974theory,nayani2015spontaneous}, leading to defect-mediated ordered structures, which are useful in designing optical microstructures\cite{stebe2009oriented}. Generally, the system is closed with static boundaries. However, the static boundary is not necessary for the systems in nature. For example, the swelling or shrinking of the outer biomembrane strongly influences the ordering inside living cells\cite{spellings2015shape} or viruses\cite{speir2012nucleic}. The effect of such a periodically moving boundary on self-assembly is still largely unknown.  Especially, despite the appearance of moving defects in active nematics \cite{shi2013topological}, it  is still unclear whether the topological defects are stable under the influence of moving boundary  and how they affect and mediate the non-equilibrium patterns of passive nematics.

 With the periodic motion of boundary, the system is dissipative, such as fluid\cite{zhang2000periodic} and granular material\cite{strassburger2000crystallization,aumaitre2003granular}. Collective motion appears and pattern formation is formed far from equilibrium. It is noted that the boundary motion herein becomes a way to input energy into the system, and the dissipation of the energy could induce the emergence of ordered structures. The generation of order resulting from the collective motion has been observed in many other dynamic self-assembly systems \cite{grzybowski2000dynamic,tagliazucchi2014dissipative,aumaitre2003granular,Schneider1986Molecules,Karsenti2008Self,zhang2000periodic,strassburger2000crystallization}.
Currently, understanding the complex collective motions  mainly depends on continuum theories. But they are invalid in systems where the effects  of particles themselves do become important. Thus, computer simulation is the most economical and effective way to address this problem and provide guidance for experimental research.

In this paper, we investigate 2D dynamic self-assembly of ellipsoids in circular confinement under a boundary with periodically stretching and contracting oscillation by Langevin dynamics simulations. Our purposes are to clarify the difference on the confinement between the dynamic and static boundary, to explore the mechanism of the dynamic self-assembly in  confined ellipsoids, and to understand the interplay between the structure and dynamics. We also note that for condensed LCs, glass transition is hard to avoid due to the limited rotational mobility. Here we expect that the input energy, aroused by the work done by the moving boundary into the system, can overcome the barrier of glass formation, leading to the ordered structures in condensed cases. Moreover, we try to verify that the boundary motion can be proved as an effective way to achieve novel steady structures in confined LCs.

\section{Models}

In the modeling, the dynamic boundary is designed as a circular boundary composed of spherical beads. To highlight the role of boundary motion, we pay more attention to strongly confined conditions. The boundary is composed of 157 beads of diameter $\sigma_{0}$ with the radius of $r_{max}=25 \sigma_{0}$. When the boundary motion is taken into account, the radius of the circle is written as $r=(r_{max}-A)+A \cos (2\pi f t)$, where A is the amplitude and f is the frequency of the boundary motion. The dimension of confined ellipsoids is chosen as $4.4 \times 1 \times 1$ $\sigma_{0}^3$ with the aspect ratio $\kappa=4.4$. To mimic a 2D system,  the height of Z direction is fixed as $\sigma_{0}$, so all the ellipsoids are confined in the X-Y plane. The number of ellipsoids within the monolayer is fixed as $N=392$ corresponding to an area fraction $\phi\approx70\%$ at $r=r_{max}$.    Additionally, the effects of system size and particle shape are studied. The interaction between all particles is adopted as Gay-Berne (GB) potential\cite{Morenorazo2011Effects,Morenorazo2012Liquid,Ji2009Structure,Bates1999Computer,berardi2007computer}, which is a short-range steric interaction resulting from a generalization of 12-6 Lennard-Jones interaction potential to treat anisotropic molecules. For simplicity, the boundary is taken as maintaining a circular shape during the course of this movement, so the interactions between boundary beads are neglected. All the boundary beads move together, which is reflected by the periodic variation of the circular radius. Our simulations are performed using LAMMPS software based on the overdamped Langevin dynamics. All the calculations are started with a random distribution of ellipsoids (see Supplementary Methods for more details).

\section{Discussion}\label{sec:3}

\subsection{Novel steady structures induced by the dynamic boundary}
\begin{figure*}
\centering
 \includegraphics[width=14cm]{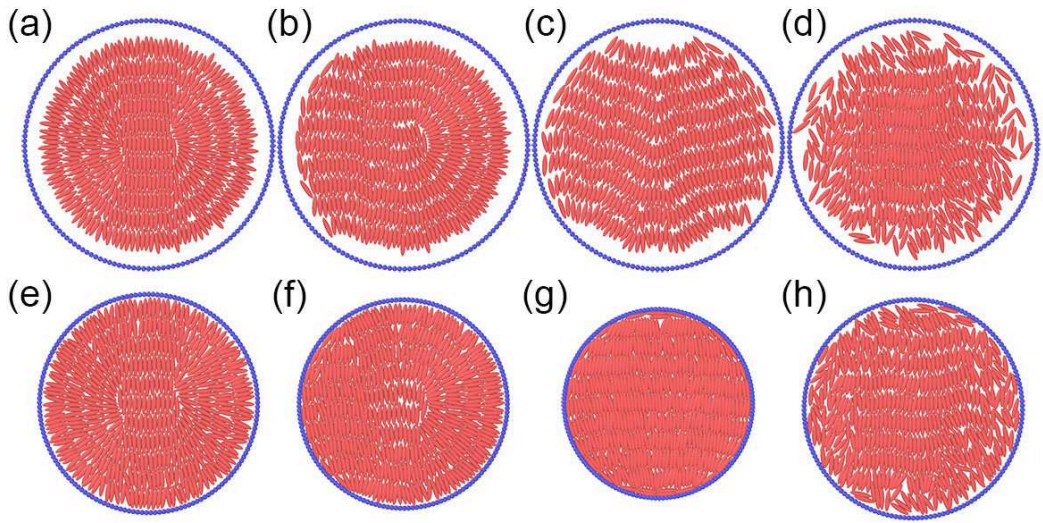}
  \caption{Pattern formation for steady structures of the confined ellipsoids driven by the boundary oscillation. The above row (a)-(d) shows the structures at $r=r_{max}$, and the under row (e)-(h) represents the structures at $r=r_{min}=r_{max}-2A$. From (a) to (c) (or (e) to (g)), the frequency of the boundary oscillation is fixed at 2.0, and the amplitude is 1.5, 2.0, and 3.0, respectively. Structures of (a)$\&$(e), (b)$\&$(f) and (c)$\&$(g) are named as \textbf{S}$_{2}$, \textbf{S}$_{1}$ and \textbf{S}$_{0}$ for simplicity. (d)$\&$(h) is named as the core-shell structure with $f=5.0$ and $A=1.5$.}
\end{figure*}
Based on our simulations, the  difference is clearly found between the confinement with static and dynamic boundaries. For the static case (Supplementary FIG. S2), the ellipsoids near the boundary are all well aligned with it. With the increase of particle volume fraction, nematic-smectic transition occurs among the internal particles\cite{Bates1999Computer}. With further increased particle density,  the motion of ellipsoidal particles is greatly hindered due to the anisotropic geometry, which frustrates the smectic phase and leads to an orientational glass, in agreement with experimental observations\cite{zheng2014structural}. However, once the boundary starts to stretch and contract periodically, the situation is completely different. Although the volume of the confined system varies periodically with the boundary oscillation, the self-organized structure of confined ellipsoids remains almost unchanged, and it is named a steady structure.  FIG. 1 (a)-(d) provides the morphology of typical steady structures at $r=r_{max}$, and (e)-(h) are those at $r=r_{min}$. When the frequency f is not large, with increasing amplitude A, different smectic structures with both positional and orientational order are achieved, shown in FIG. 1(a)-(c) and (e)-(g). Specifically, FIG. 1 (a) and (e) present concentric circles with the center of lamellar structure, having two stable defects of charge  $k_d=+1/2$. Here, the charge of topological defect $k_d$ is calculated from $\oint \frac{d\theta}{ds}ds = 2\pi k_d$, where $\theta$ is the angle between the long axis of the ellipsoid and the coordinate axis and s is the distance on the integral loop\cite{book}. FIG. 1(b) and (f) show the mixture of semicirclar and lamellar structures, having one stable defect of charge  $1/2$. FIG. 1 (c) and (g) are defect-free lamellar structure. We denote these three ordered structures as \textbf{S}$_{2}$, \textbf{S}$_{1}$, and \textbf{S}$_{0}$, respectively. Under the circular confinement, the concentric configuration is expected in dense LCs for effective packing, except for the center due to the large curvature, so that two defects are stabilized in \textbf{S}$_{2}$.  With increasing amplitude, not only the particles close to the boundary but also the inner particles participate in the collective motion. The strengthened collective motion of particles will lead to reduction and disappearance of defects. So \textbf{S}$_{1}$ and \textbf{S}$_{0}$ appear. This is very different from the equilibrium condition with the conservation of topological charge in confined LCs\cite{book}, where interactions determine the final structure rather than the collective motion of particles.  Additionally, the particle densities in these steady structures have reached or exceeded those in the glass state when the boundary is static, implying that the input energy can help the ellipsoids overcome the barrier of glass state. When the frequency is high (FIG. 1 (d) and (h)), the core-shell structure (\textbf{S}$\rm {_{cs}}$) appears, which is composed of an ordered core and a disordered shell. In this case, frequent collisions between the boundary and particles break the collective motion of particles close to the boundary, while the ordered core due to the collective motion of inner particles still exists.

\subsection{Structural diagram and structural transition}

In order to systematically study the influence of boundary motion on the self-organized structure, the structural diagram as functions of amplitude and frequency of the boundary motion is given in FIG. 2 (a). When the boundary starts to oscillate with a small amplitude and a small frequency, nematic-smectic transition occurs because particles tend to aggregate inside under the influence of boundary motion. Further increasing of A or f may lead to the appearance of out-of-equilibrium steady structures, in which the collective motion of confined ellipsoids plays the dominating role.  With larger amplitude or frequency, disordered state may occur because the boundary oscillation is too intense, the collective motion of particles is broken. It should be mentioned that the effects of amplitude and frequency on the structure are different, due to their different influences on particles collective motion.  The diagram shows the  structures from near-equilibrium to non-equilibrium, fully demonstrating the influence of the boundary motion on the self-assembly. Here, near-equilibrium structures represent those structures appearing in the equilibrium system confined with a static boundary, while non-equilibrium structures are not available in equilibrium.

\begin{figure}
\centering
 \includegraphics[width=7cm]{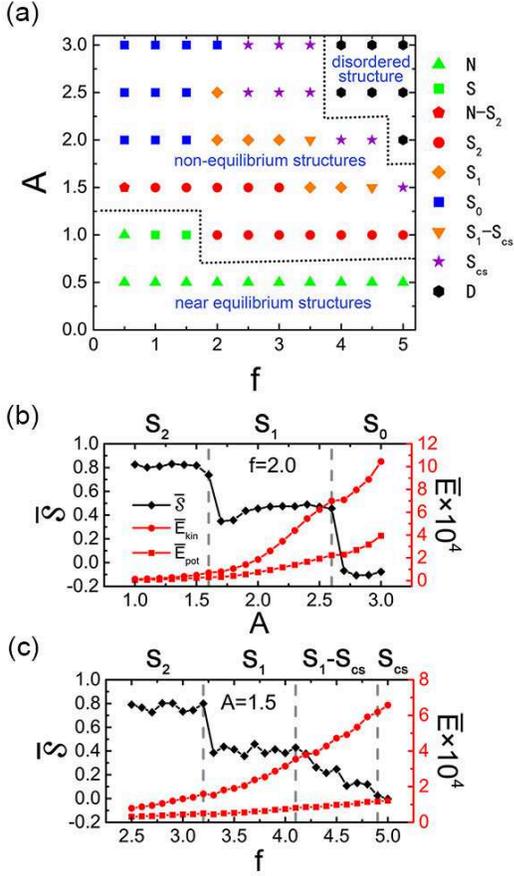}
  \caption {The structural diagram in A-f plane of the strong confinement system (a), including near equilibrium structures (\textbf{N} and \textbf{S} representing the nematic and smectic phases), non-equilibrium (\textbf{S}$_{2}$, \textbf{S}$_{1}$, \textbf{S}$_{0}$ and \textbf{S}$\rm {_{cs}}$) structures and crossovers between two structures ( \textbf{N}-\textbf{S}$_{2}$ and \textbf{S}$_{1}$-\textbf{S}$\rm {_{cs}}$). \textbf{D} is the disordered structure. Dot lines are added for the better visualization of the structural diagram. Average radical orientational order parameter $\mathcal{S}$ and average kinetic and potential energy as functions of A and f.  (b) Structural  transition driven by the change of amplitude at a fixed frequency f=2.0, and (c) transition driven by the change of frequency at a fixed amplitude A=1.5.}
\end{figure}

To quantify the structural transition, a radical orientational order parameter $\mathcal{S}$\cite{duclos2017topological,dzubiella2000topological} is introduced as, $S(t)=\frac{1}{TN}\int_{t}^{t+T}\displaystyle \mathrm d\tau \sum_{i=1}^{N}[2(\hat{\vec{u}}_{i}(\tau)\cdot\hat{\vec{r}}_{i}(\tau))^{2}-1]$. Here the unit vector $\hat{\vec{r}}_{i}$  points from the area center to the particle center. $T$ represents the period of boundary motion, and N is the number of confined ellipsoids. When $\mathcal{S}$ is positive, it is a radical oriented system. If $\mathcal{S}$ is negative, it means a system without radical aligning order. FIG. 2 (b) and (c) show the average radical orientational order parameter $\mathcal {\overline{S}}$, $\overline{E_{kin}}$ and $\overline{E_{pot}}$, by averaging one thousand cycles of boundary oscillation after the steady structure is achieved. With the increase of the amplitude, the structure undergoes \textbf{S}$_{2}$ $\to$ \textbf{S}$_{1}$ $\to$ \textbf{S}$_{0}$ in FIG. 2 (b), where $\mathcal {\overline{S}}$  shows abrupt jumps. A large value of $\mathcal {\overline{S}}$ in the condition of \textbf{S}$_{2}$ is owing to the orientation of most particles along the radial direction. A small $\mathcal {\overline{S}}$ for \textbf{S}$_{0}$ is because lots of the ellipsoids are perpendicular to the radial direction. And an intermediate $\mathcal {\overline{S}}$ is for \textbf{S}$_{1}$. The abrupt changes imply discontinuity of the three structures on $\mathcal {\overline{S}}$. As the frequency is increased in FIG. 2 (c), the structure changes from \textbf{S}$_{2}$ $\to$ \textbf{S}$_{1}$ $\to$ \textbf{S}$\rm {_{cs}}$. Abrupt change still can be seen in the transition of  \textbf{S}$_{2}$ $\to$ \textbf{S}$_{1}$, while the transition of \textbf{S}$_{1}$ $\to$ \textbf{S}$\rm {_{cs}}$ is more continuous owing to the appearance of crossover between \textbf{S}$_{1}$ and \textbf{S}$\rm {_{cs}}$. Average kinetic and potential energies are given in FIG. 2 (b) and (c).  Both of them show continuous increase, and the increase of kinetic energy is obviously larger than that of potential energy. As mentioned above, the strengthened collective motion of particles will lead to structural transitions among ordered steady structures with the decrease of defect number, which are reflected by the abrupt changes of  $\mathcal {\overline{S}}$. Here the abrupt change on $\mathcal {\overline{S}}$ and continuous variation of energy can be largely attributed to confinement. In order to study the size effect on the structural transition, we investigate two other systems, which have larger volumes and the same density of ellipsoids (Supplementary FIG. S3). The abrupt change of $\mathcal {\overline{S}}$ still can be seen, since the transition of \textbf{S}$_{2}$ $\to$ \textbf{S}$_{1}$ $\to$ \textbf{S}$_{0}$ still exists. However, the transition shifts to cases with larger amplitude, and  structures with frustrations appear in conditions with small amplitude. It can be expected that, for those systems with much larger volumes, the influence of confinement will be weakened significantly. So that ordered steady structures, including \textbf{S}$_{2}$, \textbf{S}$_{1}$ and \textbf{S}$_{0}$, are becoming more and more difficult to form. Instead, \textbf{S}$\rm {_{cs}}$ is more likely to occur when the input energy is large enough. In addition, we have studied the effect of ellipsoids with different aspect rations in strong confinement conditions (Supplementary FIG. S4).  \textbf{S}$_{2}$, \textbf{S}$_{1}$, \textbf{S}$_{0}$ and \textbf{S}$\rm {_{cs}}$ still are four typical non-equilibrium steady structures existing in different cases.

\subsection{Energies and density profiles of steady structures}

\begin{figure*}
  \centering
   \includegraphics[width=14cm]{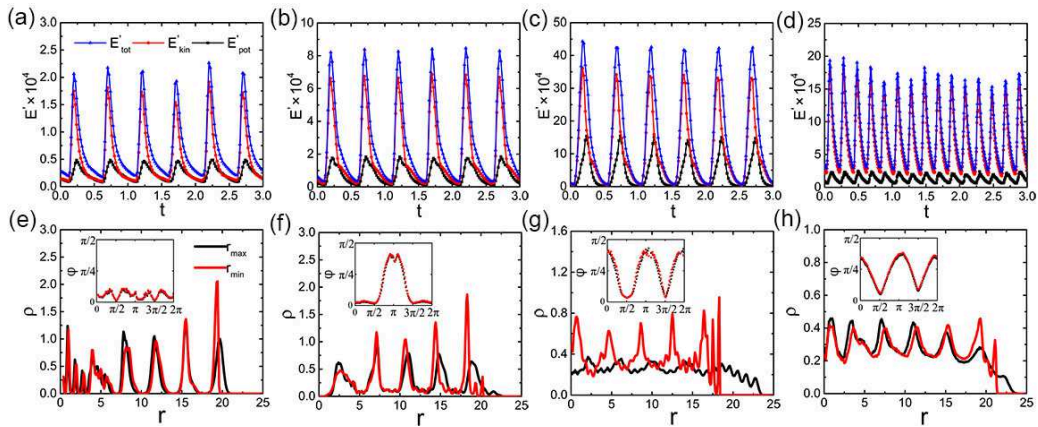}
   \caption{Energies of the system as a function of time (a)-(d), density profiles and average orientation (inset) of  particles (e)-(h) for the four steady states shown in Figure 1. The kinetic energy is $E'_{kin}(t)=\sum_{i=1}^{N}(\frac{1}{2}m v_{i}^{2}(t)+\frac{1}{2}I\omega_{i}^{2}(t)),$  where $i$ represents the i$th$ particle, the potential energy $E'_{pot}(t)$ originates in the GB potential of the system, and the total energy is the sum of kinetic energy and potential energy. Black, red and blue lines represent potential, kinetic and total energy, respectively.  The density distribution is calculated from $\rho(r)=\frac{\langle n\rangle}{2\pi r\Delta r}$, where $n$ denotes the number of particles in the annulus of $r\rightarrow r+\Delta r$. To characterize the orientation of  particles, the circular area is divided into 100 circular sectors, and the radial unit vector for each sector is defined. Then the average orientation of particles in each sector is calculated. The angle between the average orientation of the particles and the radial unit vector is Y-axis of the insets, which varies from 0 to $\frac{\pi}{2}$, with 0 corresponding to particles aligned radically and  $\frac{\pi}{2}$ for particles aligned to the boundary\cite{garlea2016finite}. The X-axis coordinate of the insets is the polar angle of the confined cavity. Black and red curves represent the cases of $r=r_{max}$ and   $r=r_{min}$, respectively. }
\end{figure*}

 To make clear the non-equilibrium nature and disclose the major factor maintaining the steady structure from the energy perspective, we show potential energy, kinetic energy and total energy for four steady structures in FIG. 3 (a)-(d). Two common features can be found. One is that all the curves show periodical oscillation. This is because the work is done to the system when boundary shrinks, and the system works externally when boundary expands. So all the energy curves go up and down. The other is that kinetic energy plays a dominate role in maintaining the steady structure. It implies that most of the work performed on the system is converted to the kinetic energy arousing the collective motion of ellipsoids.  From (a) to (c), the maximum of kinetic energy increases greatly, indicating the strengthened collective motion of confined particles with the increase of amplitude of the boundary motion. In all the four cases, the frequency of energy oscillations is the same to the frequency of boundary motion, reflecting the characteristics of the system driven by the boundary oscillation.

Given that the volume of the system is changed under the influence of boundary motion, a certain adjustment of particle distribution is required to keep the steady structure. To quantify it, the density profile and average orientation of particles\cite{garlea2016finite} are calculated in FIG. 3 (e)-(h). In FIG. 3 (e), the main difference of density profiles between $r=r_{max}$ and $r=r_{min}$ is the position and the height of the last peak. It implies that only the particles near the boundary have obvious positional change in \textbf{S}$_{2}$. With increasing amplitude, more and more particles are driven to move. Therefore, an increasing difference in density profiles between $r=r_{max}$ and $r=r_{min}$ appears in \textbf{S}$_{1}$, shown in FIG. 3 (f). The difference reaches the maximum in \textbf{S}$_{0}$, where distributions of all the particles vary with the boundary motion, as FIG. 3 (g) shows.  Once the frequency is increased and the core-shell structure forms, only the distribution of particles in the shell changes, shown in FIG. 3 (h). Although locations of particles vary during the boundary oscillation, the number of main peaks of density profiles in FIG. 3 (e)-(h) is unchanged, indicating that the positional order of structures is maintained. Furthermore, the averaged orientation of particles, reflecting the angle between the long axis of the ellipsoid and the radial direction of the round plane, is given in insets of FIG. 3 (e)-(h). In all the four cases, two curves are almost identical, implying that the orientational order of structures is well kept during the boundary oscillation.  Hence, smectic steady structures with both positional and orientational order of confined ellipsoids can be achieved.

\subsection{Collective motion of confined ellipsoids in ordered steady structures}

It is worth emphasizing that the ordered steady structures result from the collective motion of particles. In order to clearly reflect the collective movement of particles, we present average translational and rotational velocity distributions of ellipsoids for \textbf{S}$_{2}$, \textbf{S}$_{1}$ and \textbf{S}$_{0}$ at r = $r_{max}$ and $r_{min}$ separately. From FIG. 4, it can be seen that the main way of particles collective motion is the translational motion. Different velocities along the radial direction indicate a spatial inhomogeneity of particle motion. In FIG. 4 (a), when the confined volume reaches the maximum, particles close to the boundary have low speed, and those distributed in the center keep motionless. However, when the volume shrinks to the minimum (FIG. 4 (b)), outside particles gain increased velocities while inside ones still remain immovable. This characteristic of particle collective motion contributes to the formation of two defects. With increasing amplitude, although the difference in motion between outside and inside particles still exists,  the motion of inner particles is significantly increased. This will lead to the instability of topological defects. Therefore, the reduction (FIG. 4 (c) and (d)) and disappearance (FIG. 4 (e) and (f)) of defects happen in the cases with larger amplitudes. It can be concluded that two requirements are needed for the formation of stable defects. One is the inhomogeneous particle motion, and the other is the low velocity of inner particles.
\begin{figure}
 \centering
  \includegraphics[width=8cm]{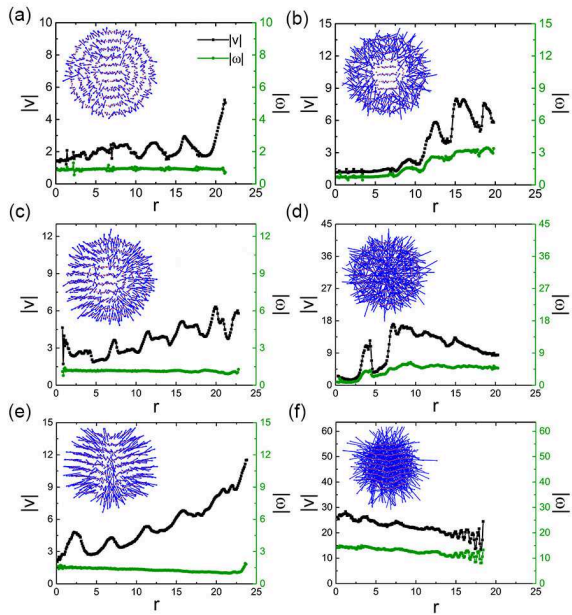}
    \caption {Average translational and rotational velocity distribution for the steady structure of \textbf{S}$_{2}$ (a) and (b), \textbf{S}$_{1}$ (c) and (d), \textbf{S}$_{0}$ (e) and (f). (a)(c)(e) are the cases with r = $r_{max}$. (b)(d)(f) are those at r = $r_{min}$. The translational (black curves) and rotational velocities (green curves) are calculated from  $\arrowvert v \arrowvert =\langle \frac{1}{n}\displaystyle\sum_{i=1}^{n}|v_{i}|\rangle$ and $\arrowvert \omega \arrowvert =\langle \frac{1}{n}\displaystyle\sum_{i=1}^{n}|\omega_{i}|\rangle$, respectively. Here $n$ denotes the number of particles in the $r\rightarrow r+\Delta r$ and $\langle\cdot\cdot\cdot\rangle$ represents the ensemble average. Insets are translational velocity distributions in 2D, where red dots denote the centroid of ellipsoids; arrows represent the direction of velocity; and blue lines reflect the magnitude of velocity.}
 \end{figure}

\subsection{Slow and fast dynamics}

Exploring dynamics of the self-assembly is another important issue. Dynamic evolutions of four steady structures are given in Movie (a)-(d) in Appendix, where the collective motion of confined ellipsoids driven by the moving boundary is notable. Taking \textbf{S}$_{2}$ and \textbf{S}$_{0}$ for examples, we provide the radical orientational order parameter and system energies as a function of time in FIG. 5. Two types of dynamics can be distinguished: one is the case (a), where the final structure is achieved after structural rearrangement and energetic adjustment. It takes some time to form \textbf{S}$_{2}$. The other is the case  (b), where  the formation of \textbf{S}$_{0}$ is very quickly, and no structural rearrangement and energetic adjustment are needed. We hereby name the former as slow dynamics, and the latter is fast dynamics. Considering that these two cases have different amplitudes, the input energy, proportional to the work performed on the system, is different. The interplay between structure and dynamics is determined by the input energy. Specifically, the larger the input energy, the faster the dynamics is, and the quickly the steady structure will be achieved.

\begin{figure}
 \centering
   \includegraphics[width=8cm]{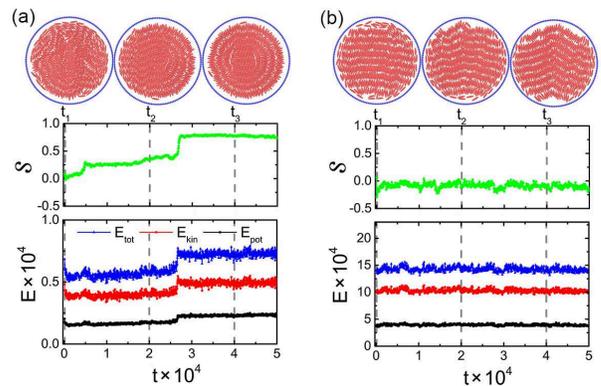}
    \caption{Snapshots of intermediate structures, radical orientational order parameter $S$, potential energy $E_{pot}$, kinetic energy $E_{kin}$, and total energy $E_{tot}$ $(E_{tot}=E_{pot}+E_{kin})$ of the system as functions of time for the slow dynamics of \textbf{S}$_{2}$ (a) and the fast dynamics of \textbf{S}$_{0}$ (b). Snapshots of intermediate structures from left to right in each case correspond to $t=t_1$, $t_2$, and $t_3$, respectively.}
 \end{figure}

\begin{figure}
  \centering
    \includegraphics[width=5.5cm]{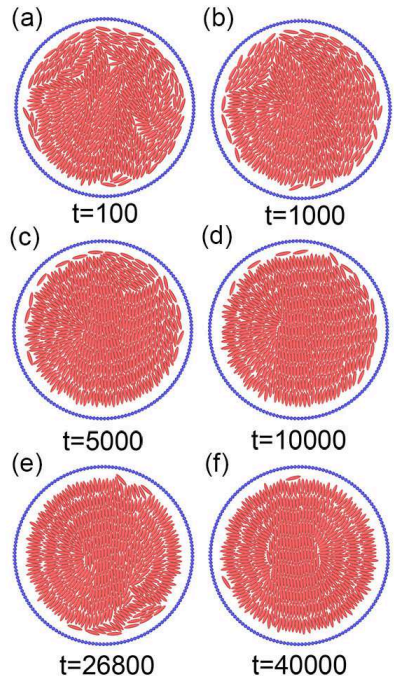}
   \caption{Snapshots of intermediate structures at $r=r_{max}$ for six different times during the formation of \textbf{S}$_{2}$. }
  \end{figure}

However, in the slow dynamics,  structural reorganization is inevitable. Intermediate structures during the formation of \textbf{S}$_{2}$ are given in FIG. 6.  At the beginning, particles close to the boundary tend to aggregate inside and multi-domain structure forms. As time goes on, the input energy is gradually transmitted to the whole system through collisions between particles. Meanwhile, orientations of neighboring domains will gradually become uniform. So the multi-domain structure will be replaced by the structure with defect. The goal of structural reorganization is to get low energy dissipation.  Although a smectic structure with one defect occurs at t=10000, structural evolution continues.  The final steady structure is the smectic structure with two stable $+1/2$ defects since the system reaches the maximum energy and minimum dissipation.

\section{Conclusions}
In conclusion, we use Langevin dynamics simulations to investigate the influence of a dynamic boundary on the self-assembly of circularly confined LCs. Non-equilibrium steady structures are achieved, which show novel smectic structures with the controllable topological defects. The collective motion of the ellipsoids, determined by the input energy due to the boundary motion, plays a vital role in determining both the structure and dynamics. Although it is a simple model, our results provide a good example of the self-assembly from equilibrium to far-from equilibrium, and disclosing the distinct difference of the confined system with the static boundary and the oscillating boundary. Moreover, the structure with two $+1/2$ defects and the core-shell structure have been observed in other similar systems, such as confined populations of spindle-shaped cells\cite{duclos2017topological} and bacteria\cite{rein2016collective}, which may reveal the generality of the formed steady structures in different non-equilibrium systems.

Our model is a simple example to show the influence of the dynamic boundary on self-assembly of colloidal particles. Since the steady structures are very sensitive to the input energy, multiple factors would affect the dynamic self-assembly, including different types of boundary motion, a changeable shape of the boundary, and different interactions between the confined particles and the boundary. Further work will pay more attention to these factors as well as other possible phenomena aroused by dynamic boundaries, such as phase separation. Another important aspect is that testing our predictions requires more experimental work. On one hand, using this strategy may help us to implement dynamic ordered steady structures in LCs. On the other hand, our results promote us to reconsider reasons of self-assembly in soft closed complex system. Does effective interaction, resulting from the movement of the boundary, lead to a condensed and ordered structure?
\begin{acknowledgements}
This work is supported by the National Natural Science Foundation of China (No. 91427302, 11474155, 11774146, and 11774147), and the Fundamental Research Funds for the Central Universities (No. 020414380045). We are grateful to the High Performance Computing Center (HPCC) of Nanjing University for doing the numerical calculations in this paper on its blade cluster system. C. Ren thanks Dr. K. Yang and Dr. W. Tian  from Soochow University for fruitful discussions.
\end{acknowledgements}

\clearpage

\end{document}